\documentclass[showpacs]{revtex4}
\usepackage{indentfirst}
\begin{document}
\title{\bf Molecular state $\Sigma_b \Sigma_b^*$ in the coupled-channel formalism}
\author{S.M. Gerasyuta}
\email{gerasyuta@SG6488.spb.edu}
\author{E.E. Matskevich}
\email{matskev@pobox.spbu.ru}
\affiliation{Department of Physics, St. Petersburg State Forest Technical
University, Institutski Per. 5, St. Petersburg 194021, Russia}
\begin{abstract}
In the framework of the dispersion relation technique the relativistic
six-quark equations for the molecule $\Sigma_b \Sigma_b^*$ are found.
The relativistic six-quark amplitudes of the hexaquark including the
quarks of three flavors ($u$, $d$, $b$) are calculated. The pole
of these amplitudes determines the mass of $\Sigma_b \Sigma_b^*$ state
$M=11620\, MeV$. The binding energy is equal to $27\, MeV$.
\end{abstract}
\maketitle
\section{Introduction.}
The prediction of a relatively deeply bound system with the quantum
numbers of $\Lambda\Lambda$ (called the H dibaryon) by Jaffe \cite{1}
in the late 1970s, based upon a bag-model calculation, started a vigorous
search for such a system, both experimentally and also with alternate
theoretical tools \cite{2, 3, 4}.

The first study of baryon-baryon interactions with the lattice QCD
was performed more than a decade ago \cite{5, 6}. This calculation
was quenched and used the $m_{\pi}\ge 550\, MeV$. The NPLQCD Collaboration
performed the first $n_f=2+1$ QCD calculations of baryon-baryon
interactions \cite{7, 8} at low energies but at unphysical pion masses.
Quenched and dynamical calculations were subsequently performed by the
HAL QCD Collaborations \cite{9, 10}.

A number of quenched lattice QCD calculations \cite{11, 12,13, 14, 15, 16}
have considered the H dibaryon, but to  date no definite results have
been reported. Earlier work concluded that the H dibaryons does not
exist as a stable hadron in quenched QCD \cite{15}, while more recent
works \cite{16, 17} find a hint of a bound state. By using energy
dependent potential in the Schredinger equation in $SU(3)$ limit,
a hint of H dibaryon has been found in Ref. \cite{18}. However, this
hint evaporates when $SU(3)$ breaking is included \cite{19}.

The results of the lattice QCD calculations are presented in Ref. \cite{20}.
The first clear evidence for a bound state of two $\Lambda$ baryons directly
from QCD at a pion mass of $m_{\pi}\sim 389\, MeV$ is provided.

In our previous paper \cite{21} the relativistic six-quark equations are found
in the framework of coupled-channel formalism. The dynamical mixing between
the subamplitudes of hexaquark are considered. The six-quark amplitudes
of dibaryons are calculated. The poles of these amplitudes determine the
masses of dibaryons.

The relativistic six-quark equations for the molecule $N\Xi$ are found in
the framework of the dispersion relation technique. The relativistic
six-quark amplitudes of the hexaquark including the quarks of three
flavors ($u$, $d$, $s$) are calculated. The contributions of six-quark
subamplitudes to the hexaquark amplitude are calculated. The pole of these
amplitudes determines the mass of dibaryon state $N\Xi$ $M=2252\, MeV$.
The binding energy is equal to $3\, MeV$.

In the framework of the dispersion relation technique the relativistic
six-quark equations for the bottom baryon molecule $\Sigma_b \Sigma_b^*$
are found \cite{22}. The relativistic six-quark amplitudes of the hexaquark
including the two heavy quarks are calculated. The pole of these amplitudes
determines the bottom molecule $\Sigma_b \Sigma_b^*$ $(uub)(uub)$ with the
mass $M=11620\, MeV$. The binding energy is equal to $27\, MeV$.

\section{Six-quark amplitudes of molecular state $\Sigma_b \Sigma_b^*$.}

The relativistic six-quark equations in the framework of the dispersion
relation technique are derived. Only planar diagrams are used. The other
diagrams due to the rules of $1/N_c$ expansion \cite{23, 24, 25} are neglected.
We shall consider the derivation of the relativistic generalization
of the Faddeev-Yakubovsky approach \cite{26}. In our case the bottom
dibaryons with the two bottom quarks are used. The pairwise interaction
of all six quarks in the hexaquark are considered.

For instance, we consider the reduced amplitude $\alpha_2^{1^{uu}1^{uu}}$
(Fig. 1). The coefficients are determined by the permutation of quarks
\cite{27, 28}. We should use the coefficients multiplying of the
diagrams in the graphical equation (Fig. 1).

In Fig. 1 the first coefficient is equal to 4, that the number $4=2$
(permutation of particles 1 and 2) $\times 2$ (permutation of particles
3 and 4); the second coefficient is equal to 8, that the number $8=2$
(permutation of particles 1 and 2) $\times 2$ (replacement of particles
1, 2 with 3, 4) $\times 2$ (permutation of particles
5 and 6); the third coefficient is equal to 16, that the number $16=2$
(permutation of particles 1 and 2) $\times 2$ (permutation of particles
3 and 4) $\times 2$ (replacement of particles 1, 2 with 3, 4) $\times 2$
(permutation of particles 5 and 6).

The system of equations for the molecule $\Sigma_b \Sigma_b^*$ is given
in the Appendix A. The functions $I_1$, $I_2$, $I_3$, $I_4$, $I_5$, $I_6$,
$I_7$ similar to the Ref. \cite{22} are considered.

\section{Calculation results.}

The system of five equations allows us to consider the bottom molecule
$\Sigma_b \Sigma_b^*$ $(uub\, uub)$ with the bottom $B=-2$, the isospin
$I=2$ and the spin-parity $J^P=2^+$. The mass of this state is equal to
$M=11620\, MeV$. The binding energy is determined by the energy
threshold $11647\, MeV$ \cite{29}.

The model in question take into account the hexaquarks with the two
heavy quarks $qqqqQQ$, $q=u, d$, $Q=b$. In these calculations we do
not consider the hexaquarks with the two different heavy quarks. But
the results will be similar to the present ones.

The experimental data are absent, therefore we use the dimensionless
parameters, which are similar to the previous paper \cite{22}. The quark
masses of the model are $m_{u,d}=495\, MeV$ and $m_b=4840\, MeV$.
We use the gluon coupling constants $g_0=0.653$ (diquark $0^+$) and $g_1=0.292$
(diquark $1^+$), cutoff parameters $\Lambda=11$, $\Lambda_{qb,bb}=7.35$.
The choice of cutoff values is reasonable in our consideration.
If the cutoff is decreased, the hexaquarks masses will be larger.

In quark models, which describe rather well the masses and static properties
of hadrons, the masses of the quarks usually have the similar values for the
spectra of light and heavy hadrons. However, this is achieved at the expence
of some difference in characteristic of confinement potential. It should be
borne in mind that for a fixed hadron masses of the constituent quarks
which enter into the composition of the hadron will become smaller when
the slope of the confinement potential increases or its radius decreases.
Therefore, conversely, we can change the masses of the constituent quarks
when going from the spectrum of light to the heavy hadrons, while keeping
the characteristic of the confinement potential unchanged.

We can effectively take into account the contribution of the confinement potential
in obtaining the spectrum of heavy hadrons. The masses distinction of $u$ and $d$
quarks is neglected. The estimations of the theoretical error on the heavy
dibaryons masses is $1\, MeV$. This results was obtained by the choice of
model parameters.

\section{Conclusion.}

In a strongly bound systems, which include the light quarks, where
$p/m \sim 1$, the approximation of nonrelativistic kinematics and dynamics
is not justified. In our paper, the relativistic description of six-particles
amplitudes of heavy dibaryons with the two heavy quarks is considered. We take
into account the $u$, $d$, $b$ quarks. Our model is confined to the
quark-antiquark pair production on account of the phase space restriction.
Here $m_q$ and $m_Q$ are the "masses" of the constituent quarks. Therefore the
production of new quark-antiquark pair is absent for the low-lying hadrons.

Hadronic molecules are loosely bound states of hadrons, whose inter-hadron
distances are larger than the quark confinement size.

The discovery of $Z_b$ states raises an interesting possibility of a strongly
bound $\Sigma_b^+\Sigma_b^-$ deuteron-like state, a "beauteron" \cite{30}.
But the results of our calculations showed that the mass of bound state
$\Sigma_b^+\Sigma_b^-$ dibaryon is equal to $M=10290\, MeV$. The energy
threshold of this state is equal to $11627\, MeV$ \cite{29}. The binding energy
is very large, therefore this state is not hadronic molecule. We predict
only one bottom dibaryon molecule $\Sigma_b \Sigma_b^*$ $(uub\, uub)$ with
the following quantum numbers: the isospin $I=2$, the spin-parity $J^P=2^+$
and the bottom number $B=-2$. The mass of this state is equal to
$M=11620\, MeV$. The binding energy is determined by the energy
threshold ($E_B=27\, MeV$).

The loosely bound state does not exist for the $\Lambda_b \Lambda_b$ system.
The bottom baryon molecule $\Sigma_b \Sigma_b^*$ may be produced at LHC.
It should also be seen in lattice QCD.

\begin{acknowledgments}
The work was carried with the support of the Russian Ministry of Education
(grant 2.1.1.68.26) and RFBR, Research Project No. 13-02-91154.
\end{acknowledgments}

\newpage

\appendix

\section{The system of equations for the bottom molecule $\Sigma_b \Sigma_b^*$.}

\begin{eqnarray}
\label{A1}
\alpha_1^{1^{uu}}&=&\lambda+4\, \alpha_1^{1^{uu}} I_1(1^{uu}1^{uu})
+4\, \alpha_1^{0^{ub}} I_1(1^{uu}0^{ub})
+2\, \alpha_2^{1^{uu}1^{uu}} I_2(1^{uu}1^{uu}1^{uu})
+8\, \alpha_2^{1^{uu}0^{ub}} I_2(1^{uu}1^{uu}0^{ub})\\
&&\nonumber\\
\label{A2}
\alpha_1^{1^{bb}}&=&\lambda+8\, \alpha_1^{0^{ub}} I_1(1^{bb}0^{ub})\\
&&\nonumber\\
\label{A3}
\alpha_1^{0^{ub}}&=&\lambda+3\, \alpha_1^{1^{uu}} I_1(0^{ub}1^{uu})
+\alpha_1^{1^{bb}} I_1(0^{ub}1^{bb})
+4\, \alpha_1^{0^{ub}} I_1(0^{ub}0^{ub})
+6\, \alpha_2^{1^{uu}0^{ub}} I_2(0^{ub}1^{uu}0^{ub})\\
&&\nonumber\\
\label{A4}
\alpha_2^{1^{uu}1^{uu}}&=&\lambda+4\, \alpha_1^{1^{uu}} I_3(1^{uu}1^{uu}1^{uu})
+8\, \alpha_1^{0^{ub}} I_4(1^{uu}1^{uu}0^{ub})
+16\, \alpha_2^{1^{uu}0^{ub}} I_7(1^{uu}1^{uu}1^{uu}0^{ub})\\
&&\nonumber\\
\label{A5}
\alpha_2^{1^{uu}0^{ub}}&=&\lambda+\alpha_1^{1^{uu}} (2\, I_3(1^{uu}0^{ub}1^{uu})+I_4(0^{ub}1^{uu}1^{uu}))
+\alpha_1^{0^{ub}} (2\, I_3(1^{uu}0^{ub}0^{ub})+2\, I_4(1^{uu}0^{ub}0^{ub}))\nonumber\\
&&\nonumber\\
&+&\alpha_2^{1^{uu}0^{ub}}(2\ I_5(1^{uu}0^{ub}1^{uu}0^{ub})+2\ I_6(1^{uu}0^{ub}0^{ub}1^{uu})
+2\ I_7(1^{uu}0^{ub}1^{uu}0^{ub})+2\ I_7(1^{uu}0^{ub}0^{ub}1^{uu})\nonumber\\
&&\nonumber\\
&+&2\ I_7(0^{ub}1^{uu}1^{uu}0^{ub}))
\end{eqnarray}

We used the functions $I_1$, $I_2$, $I_3$, $I_4$, $I_5$, $I_6$, $I_7$ (Ref. \cite{22}).

\newpage

\begin{picture}(600,90)
\put(0,45){\line(1,0){18}}
\put(0,47){\line(1,0){17.5}}
\put(0,49){\line(1,0){17}}
\put(0,51){\line(1,0){17}}
\put(0,53){\line(1,0){17.5}}
\put(0,55){\line(1,0){18}}
\put(30,50){\circle{25}}
\put(19,46){\line(1,1){15}}
\put(22,41){\line(1,1){17}}
\put(27.5,38.5){\line(1,1){14}}
\put(31,63){\vector(1,1){20}}
\put(31,38){\vector(1,-1){20}}
\put(47.5,60){\circle{16}}
\put(47.5,40){\circle{16}}
\put(55,64){\vector(3,2){18}}
\put(55,36){\vector(3,-2){18}}
\put(55,64){\vector(3,-2){18}}
\put(55,36){\vector(3,2){18}}
\put(78,75){1}
\put(64,77){$u$}
\put(78,53){2}
\put(69,58){$u$}
\put(78,41){3}
\put(69,38){$u$}
\put(78,18){4}
\put(65,16){$u$}
\put(54,80){5}
\put(37,80){$b$}
\put(54,13){6}
\put(37,12){$b$}
\put(41.5,56){\small $1^{uu}$}
\put(41.5,36){\small $1^{uu}$}
\put(90,48){$=$}
\put(30,-15){$\alpha_2^{1^{uu}1^{uu}}$}
\put(110,45){\line(1,0){19}}
\put(110,47){\line(1,0){21}}
\put(110,49){\line(1,0){23}}
\put(110,51){\line(1,0){23}}
\put(110,53){\line(1,0){21}}
\put(110,55){\line(1,0){19}}
\put(140,60){\circle{16}}
\put(140,40){\circle{16}}
\put(147.5,64){\vector(3,2){18}}
\put(147.5,36){\vector(3,-2){18}}
\put(147.5,64){\vector(3,-2){18}}
\put(147.5,36){\vector(3,2){18}}
\put(128,55){\vector(1,3){11}}
\put(128,45){\vector(1,-3){11}}
\put(170,75){1}
\put(156,77){$u$}
\put(170,53){2}
\put(161,58){$u$}
\put(170,41){3}
\put(161,38){$u$}
\put(170,18){4}
\put(157,16){$u$}
\put(143,86){5}
\put(126,84){$b$}
\put(143,08){6}
\put(126,9){$b$}
\put(134,57){\small $1^{uu}$}
\put(134,37){\small $1^{uu}$}
\put(140,-15){$\lambda$}
\put(183,48){$+$}
\put(201,48){4}
\put(216,45){\line(1,0){18}}
\put(216,47){\line(1,0){17.5}}
\put(216,49){\line(1,0){17}}
\put(216,51){\line(1,0){17}}
\put(216,53){\line(1,0){17.5}}
\put(216,55){\line(1,0){18}}
\put(246,50){\circle{25}}
\put(235,46){\line(1,1){15}}
\put(238,41){\line(1,1){17}}
\put(243.5,38.5){\line(1,1){14}}
\put(247,63){\vector(1,1){20}}
\put(247,38){\vector(1,-1){20}}
\put(270,80){5}
\put(254,80){$b$}
\put(270,13){6}
\put(254,13){$b$}
\put(266,50){\circle{16}}
\put(274,50){\vector(3,2){17}}
\put(274,50){\vector(3,-2){17}}
\put(251,61){\vector(1,0){40}}
\put(251,39){\vector(1,0){40}}
\put(299,60){\circle{16}}
\put(299,40){\circle{16}}
\put(307,61){\vector(3,1){20}}
\put(307,61){\vector(3,-1){20}}
\put(307,39){\vector(3,1){20}}
\put(307,39){\vector(3,-1){20}}
\put(324,72){1}
\put(316,70){$u$}
\put(329,56){2}
\put(309,52){$u$}
\put(329,40){3}
\put(309,44){$u$}
\put(324,22){4}
\put(316,24){$u$}
\put(274,66){1}
\put(262,66){$u$}
\put(274,54){\small 2}
\put(285,51){$u$}
\put(274,40){\small 3}
\put(285,44){$u$}
\put(274,28){4}
\put(263,29){$u$}
\put(260,47){\small $1^{uu}$}
\put(293,57){\small $1^{uu}$}
\put(293,37){\small $1^{uu}$}
\put(228,-15){$4\, \alpha_1^{1^{uu}}\, I_3(1^{uu}1^{uu}1^{uu})$}
\put(343,48){$+$}
\put(359,48){8}
\put(372,45){\line(1,0){18}}
\put(372,47){\line(1,0){17.5}}
\put(372,49){\line(1,0){17}}
\put(372,51){\line(1,0){17}}
\put(372,53){\line(1,0){17.5}}
\put(372,55){\line(1,0){18}}
\put(402,50){\circle{25}}
\put(391,46){\line(1,1){15}}
\put(394,41){\line(1,1){17}}
\put(399.5,38.5){\line(1,1){14}}
\put(418,62){\circle{16}}
\put(423.5,68.5){\vector(1,1){15}}
\put(423.5,68.5){\vector(1,-1){18}}
\put(414,50){\vector(1,0){28}}
\put(450,50){\circle{16}}
\put(458,50){\vector(3,1){22}}
\put(458,50){\vector(3,-1){22}}
\put(437,59){1}
\put(430,65){$u$}
\put(431,39){2}
\put(425,52){$u$}
\put(474,62){1}
\put(464,58){$u$}
\put(474,30){2}
\put(464,36){$u$}
\put(440,85){5}
\put(427,84){$b$}
\put(418,37){\circle{16}}
\put(425,32){\vector(3,-1){20}}
\put(425,32){\vector(2,-3){12}}
\put(403,38){\vector(1,-3){8}}
\put(448,22){3}
\put(440,31){$u$}
\put(439,7){4}
\put(428,9){$u$}
\put(413,7){6}
\put(400,9){$b$}
\put(412,59){\small $0^{ub}$}
\put(444,47){\small $1^{uu}$}
\put(412,34){\small $1^{uu}$}
\put(390,-15){$8\, \alpha_1^{0^{ub}}\, I_4(1^{uu}1^{uu}0^{ub})$}
\end{picture}

\vskip60pt
\begin{picture}(600,90)
\put(90,48){$+$}
\put(107,48){16}
\put(125,45){\line(1,0){18}}
\put(125,47){\line(1,0){17.5}}
\put(125,49){\line(1,0){17}}
\put(125,51){\line(1,0){17}}
\put(125,53){\line(1,0){17.5}}
\put(125,55){\line(1,0){18}}
\put(155,50){\circle{25}}
\put(144,46){\line(1,1){15}}
\put(147,41){\line(1,1){17}}
\put(152.5,38.5){\line(1,1){14}}
\put(174,57){\circle{16}}
\put(171,37){\circle{16}}
\put(181,61){\vector(1,1){15}}
\put(156,62.5){\vector(3,1){40}}
\put(204,77){\circle{16}}
\put(181,61){\vector(1,-1){12}}
\put(179,35){\vector(1,1){14}}
\put(201,50){\circle{16}}
\put(209,50){\vector(3,2){18}}
\put(209,50){\vector(3,-2){18}}
\put(212,78){\vector(3,2){18}}
\put(212,78){\vector(3,-2){18}}
\put(179,35){\vector(1,-1){16}}
\put(156,38){\vector(1,-2){11}}
\put(235,87){1}
\put(216,88){$u$}
\put(235,64){2}
\put(222,74){$u$}
\put(231,54){3}
\put(220,50){$u$}
\put(231,32){4}
\put(216,33){$u$}
\put(185,80){1}
\put(170,75){$u$}
\put(192,63){2}
\put(180,65){$u$}
\put(182,46){3}
\put(188,57){$u$}
\put(185,32){4}
\put(190,39){$u$}
\put(198,11){5}
\put(184,12){$b$}
\put(170,9){6}
\put(156,10){$b$}
\put(168,54){\small $1^{uu}$}
\put(165,34){\small $0^{ub}$}
\put(198,74){\small $1^{uu}$}
\put(195,47){\small $1^{uu}$}
\put(130,-15){$16\, \alpha_2^{1^{uu}0^{ub}}\, I_7(1^{uu}1^{uu}1^{uu}0^{ub})$}
\put(20,-50){Fig. 1. The graphical equations of the reduced amplitude $\alpha_1^{1^{uu}}$ (\ref{A4}).}
\end{picture}

\end{document}